\documentstyle[12pt,epsf]{ioplppt}
\begin{document}
\jl{3}
\title{Metallic spin glasses}
\author{Subir Sachdev and N. Read}
\address{Department of Physics, Yale University, P.O. Box 208120,
New Haven CT 06520-8120, USA.}

\begin{abstract}
Recent work on the zero temperature phases and phase transitions of
strongly random electronic system is reviewed. The transition
between the spin glass and quantum paramagnet is examined, for 
both metallic and insulating systems. Insight gained
from the solution of infinite range models leads to a quantum field
theory for the transition between a metallic quantum paramagnetic and
a metallic spin glass. The finite temperature phase diagram is described and
crossover functions are computed in mean field theory.
A study of fluctuations about mean field leads to the formulation of scaling
hypotheses.
\end{abstract}
\maketitle

\section{Introduction}
The vast majority of the existing theoretical and experimental work on spin
glasses~\cite{hertz} has been in regimes where quantum effects can be almost
entirely neglected, and both the statics and dynamics can be described by
classical statistical mechanics models. This earlier work has focused on the
paramagnetic phase not too far above the spin glass transition temperature,
$T_c$, and on the phase with spin glass order with temperatures $T < T_c$.
Thermal fluctuations then dominate, and the typical relaxation frequency,
$\omega$, satisfies $\hbar \omega \ll k_B T$, justifying the use of classical
models. One important consequence of this classical behavior is that the
coupling between spin and charge fluctuations is rather weak; whether the 
frozen spins in the spin glass phase are in a metallic, or a 
insulating, host is not of direct importance in theories of the formation 
of the spin glass phase, or of its critical fluctuations. The metallic systems
do have a longer range RKKY interaction between the spins, but this does not
fundamentally differentiate the basic theoretical framework used to describe the two
cases.

The situation is quite different when we vary $T_c$ as a function of some
tuning parameter ({\it e.g.\/} pressure, doping concentration of spins,
or disorder) $r$,
and examine the regime where $T_c (r)$ becomes much smaller than 
a typical microscopic exchange constant~\cite{pw}. In particular, it may be
possible to  reach a $r=r_c$ at which we first have $T_c (r_c) = 0$; 
we will choose an overall shift in the value of $r$ so that $r_c = 0$.
For $r >
0$, then, the system is a paramagnet at all temperatures, including at $T=0$.
Considering the situation precisely at $T=0$, the ground state undergoes a
phase transition, from a spin glass ($r<0$) to a paramagnet ($r>0$),
which is driven entirely by quantum fluctuations and disorder. Under these
circumstances, as is described in more detail below, quantum effects play
a fundamental role over a substantial portion of the $T$, $r$ plane (and not
just at $T=0$). Concomitantly, the coupling between spin and charge
fluctuations becomes much more important. It is now essential to specify whether
the spin glass and paramagnetic phases are metallic or insulating, as the
effective theory of spin fluctuations is quite different in the two cases.

In this paper, we will review recent work on insulating and metallic quantum
spin glasses, highlighting the physical origin of the differences between the
two cases. The emphasis will be on the metallic case, primarily because it is
of greater experimental interest in the 
context of rare-earth intermetallic compounds described elsewhere in these
proceedings. A few results described here have not been explicitly obtained
before, but they are straightforward consequences of work published earlier.

\section{General considerations\label{general}}
It is useful to begin by making a 
few general remarks on the expected phases, and
their properties, of {\it strongly} random electronic systems in three
dimensions. For concreteness, consider the $T=0$ properties of the following
Hamiltonian
\begin{equation}
H = -\sum_{i < j , \alpha} t_{ij} c_{i\alpha}^{\dagger} c_{i\alpha}
- \sum_{i < j , \mu} J_{ij}^{\mu} S_{i\mu} S_{j\mu} + H_{int},
\label{hamgeneral}
\end{equation}
where $c_{i\alpha}$ annihilates an electron on site $i$ with spin
$\alpha=\uparrow, \downarrow$,
and the spin operator $S_{i\mu} \equiv \sum_{\alpha\beta} c_{i\alpha}^{\dagger}
\sigma^{\mu}_{\alpha\beta} c_{i\beta}/2$ with $\sigma^{\mu}$ the Pauli matrices.
The sites are placed in three dimensional space and labeled by
$i,j$, the hopping matrix elements $t_{ij}$
are short-ranged and possibly random, and the $J_{ij}^{\mu}$ are 
random exchange
interactions, possibly with spin anisotropies. The remainder $H_{int}$
includes other
possible short-range interactions between the electrons: we
will not need to specify them explicitly.

Figure 1 shows a possible $T=0$ phase diagram of $H$ as its couplings are
varied. We have only identified phases between which it is possible to make a
sharp distinction. The phases are determined by the behavior of their
spin and charge fluctuations. Charge transport is characterized by the $T=0$
value of the conductivity, $\sigma$; if $\sigma = 0$ the phase is an insulator,
and is metallic otherwise. For the spin sector, the uniform spin susceptibility
is not a useful diagnostic (as will become clear below), 
and we distinguish phases by whether the ground state has infinite memory of
the spin orientation on a site or not. The time-averaged moment on a given site
is denoted by $\langle S_i \rangle$ (we will generally drop vector indices,
$\mu$, $\nu$ except where needed) and it can either vanish at every site,
or take a non-zero value which varies randomly from site to site. The average
over all sites will be denoted by $\overline{\langle S_i \rangle}$, 
and will be 
non-zero only in ferromagnetic phases, which we will not consider here.

The four phases in Figure 1 are:
\newline
(1) METALLIC QUANTUM PARAMAGNET (MQP)
\newline
More simply known as the familiar `metal', this phase has
\begin{equation}
\sigma \neq 0 ~~~~~~~~~~\langle S_i \rangle = 0.
\end{equation}
The charge transport occurs through quasiparticle excitations (which also
carry spin) which can be described by a disordered version of Fermi liquid
theory~\cite{aa}: the quasiparticles still obey a transport equation as in
Landau's theory but have wave functions which are spatially disordered. 
However, it is also necessary to include a finite concentration of local 
spin moments to obtain a complete description of the low energy spin
excitations~\cite{milica}. These local moments are created by relatively weak
fluctuations in the disorder, and interact with each other and the conduction
electrons via spin exchange. Although the spin fluctuations on the local moment
sites are relatively slow, they eventually lose memory of their 
orientation, and we always have $\langle S_i \rangle = 0$.
\newline
(2) INSULATING QUANTUM PARAMAGNET (IQP)
\newline
This phase has
\begin{equation}
\sigma = 0 ~~~~~~~~~~\langle S_i \rangle = 0
\end{equation}
and is accessed by a metal-insulator transition from the MQP phase.
It is also referred to as the `random singlet' phase~\cite{bl}, the name 
describing 
spin singlet bonds between pairs of spins. Breaking any of these singlet bonds
costs an energy which has a broad distribution, with significant weight 
at low
energy. The large density of low energy spin excitations means that the spin
susceptibility is quite large, and may even diverge as $T \rightarrow 0$ (this
can also happen in the MQP phase, as argued by Bhatt and 
Fisher~\cite{milica}). This is the reason we 
have not used the spin susceptibility as a diagnostic for the phases.
\newline
(3) METALLIC SPIN GLASS (MSG)
\newline
This phase has
\begin{equation}
\sigma \neq 0 ~~~~~~~~~~\langle S_i \rangle \neq 0,
\end{equation}
and is accessed from the MQP by a spin-freezing transition. The local moments
of the MQP phase now acquire a definite orientation, and retain memory of 
this
orientation for infinite time. The Fermi liquid quasiparticle excitations are
still present and are responsible for the non-zero $\sigma$; the frozen
moments appear as random local magnetic fields to the itinerant quasiparticles.
Alternatively, we may view the spin-freezing transition as the onset, from
the MQP phase, of a spin density wave with random offsets in its phase and
orientation, as appears to be the case in recent experiments~\cite{mydosh}; this
suggests the name ``\underline{spin density glass}''. Although there are quantitative
differences  between these two points of view, there is no
sharp or qualitative distinction and the two pictures are expected
to be continuously connected. The latter point of view was explored by
Hertz~\cite{hertzsg} some time ago, but he did not focus on the vicinity of the
$T=0$ transition between the MSG and MQP phases. An infinite range model with a MSG
phase was studied in Ref~\cite{oppermann} with a static ansatz for the order
parameter. Here we will review the recent complete solution for the
MSG phase and the MSG-MQP transition in the infinite range model~\cite{sro,sg},
the Landau theory for the short range case~\cite{rsy,sro} and the consequences of
fluctuations about the mean field theory.
\newline
(4) INSULATING SPIN GLASS (ISG)
\newline
This phase has
\begin{equation}
\sigma = 0 ~~~~~~~~~~\langle S_i \rangle \neq 0
\end{equation}
Charge fluctuations are unimportant, and the collective frozen spin configuration is
expected to be well described by an effective classical spin model. Examples of
phases of this type may be found in Ref~\cite{hertz}. 

Let us emphasize that the above discussion was restricted to $T=0$, and thus
describes only the ground state properties of $H$. Indeed, all of the ground states
have been studied earlier, and we shall have little to add to their description here.
Our focus shall primarily be on the finite $T$ properties in the vicinity of
quantum phase transitions between the phases. As we will see, 
this finite $T$ behavior can be quite non-trivial and remarkably rich; our results
are summarized in Fig~\ref{f2} and will be discussed in more detail later.
The possibility that such finite $T$ crossovers may offer an explanation of the 
unusual properties of lightly doped cuprates (and perhaps other strongly 
correlated systems) was discussed in Ref~\cite{sy1}.

In the following section we will describe solutions of infinite range models of the
above phases. The insights from these solutions will be used to attack realistic
models with short-range interactions in subsequent sections.

\section{Infinite range models}
We describe three models, which address different aspects of the phase diagram in
Fig~1.
\subsection{Insulating Heisenberg spins\label{heisenberg}}
This subsection reviews the results of Ref~\cite{sy2}.

We consider the Hamiltonian $H$ with $t_{ij} = 0$ and $H_{int}$ constructed so that
the system has exactly one electron per site. 
The ground state is therefore necessarily insulating, and we are considering a model 
of the IQP and ISG phases.
The exchange
interactions $J_{ij}$ are isotropic in spin space and are independent random
variables with
$\overline{J_{ij}} = 0$ and $\overline{J_{ij}^2} = J_0^2 /N_s$, where $N_s$ is the
total number of sites. As the mean square value of $J_{ij}$ is independent of $|i-j|$,
the model has infinite range interactions, and each interaction is of order
$1/\sqrt{N_s}$.

Even this simple model is, however, not solvable. We need to consider a
generalization of the symmetry of the model from $SU(2)$ to $SU(N)$ ($N$ counts the
number of spin components of each electron), and
also consider the dependence on the ``length'' of the spin at each site 
(labeled by an
integer $n_b$). We then have a family of models labeled by the two integers $N$,
$n_b$ and the familiar spin-1/2 Heisenberg model corresponds to $N=2$, $n_b = 1$. It
has been shown that the infinite range model is at least partially solvable in the
limit
$N \rightarrow \infty$, $n_b \rightarrow \infty$ but with $n_b / N$ fixed at an
arbitrary value, and also in the limit $N \rightarrow \infty$, $n_b$ fixed.

For $N$ large, the model is in the ISG phase for $n_b / N > \kappa_c$,
where $\kappa_c$ is a numerical constant whose value is not precisely known.
In this phase we have
\begin{equation}
\lim_{\tau \rightarrow \infty}
\overline{\langle S_i (0) \cdot S_i (\tau) \rangle} = q_{EA} \neq 0
\end{equation}
where $\tau$ is imaginary time, and $q_{EA}$ the Edwards-Anderson order parameter.
The nature of the approach of the two-point spin correlator to $q_{EA}$ at large
$\tau$, and the manner in which $q_{EA}$ vanishes as $n_b / N$ approaches $\kappa_c$
from above at $T=0$ have not been determined. 

The IQP phase appears for $n_b / N < \kappa_c$, and its properties have been
determined a little more explicitly. We now have 
\begin{equation}
\overline{\langle S_i (0) \cdot S_i (\tau) \rangle} \sim \frac{1}{\tau}
\end{equation}
for large $\tau$ at $T=0$. The decay of the correlator is quite slow, and as
a result, the local dynamic linear spin susceptibility, $\chi_L ( \omega)$
diverges logarithmically as the frequency, $\omega$, approaches zero:
\begin{equation}
\chi_L ( \omega ) \sim \ln \left( \frac{1}{|\omega |} \right) + i \frac{\pi}{2}
\mbox{sgn} (\omega )
\end{equation} 
In the present model, each spin has the option of forming singlet bonds
with a very large number of partners, and the diverging local susceptibility of the
IQP phase appears to be due to strong resonance between different pairings. 

For comparison with the other infinite range models considered below, we note that
the local spectrum on each site consists of a single doubly degenerate level of a up
or a down spin, which are then coupled together by the inter site exchange
interaction. As we will see shortly, this local degeneracy is intimately linked with
the diverging local susceptibility of the present model.

\subsection{Spins in a metal\label{metal}}
This subsection reviews the results of Refs~\cite{sro,sg}.

We now consider the Hamiltonian $H$ with {\it both} $t_{ij}$ and $J_{ij}$ 
independent, Gaussian, random variables, with mean square values independent of
$i,j$. The couplings in $H_{int}$ are chosen so that the ground state 
is metallic, and therefore produces a mean field theory
of the MSG and MQP phases, and of the quantum phase transition between them.
Unlike the model of Section~\ref{heisenberg}, we have available an essentially
exact solution of the low energy properties of the present model, including 
that of the critical
properties of the MSG-MQP transition, and of the finite $T$ crossovers.

We will limit our discussion here to $T=0$; non-zero $T$ will be discussed later in
the paper. In the MQP (or disordered Fermi liquid) phase of this model we have
\begin{equation}
\overline{\langle S_i (0) \cdot S_i (\tau) \rangle} \sim \frac{1}{\tau^2}
\label{metale1}
\end{equation}
for large $\tau$. This $1/\tau^2$ decay of spin correlations is characteristic of
that found in any Fermi liquid. This is perhaps clearer in the Fourier
transform to the dynamic local susceptibility, which has the linear
form
\begin{equation}
\chi^{\prime\prime}_L ( \omega ) \sim \omega
\label{chifl}
\end{equation}
of the particle-hole continuum. As one approaches the transition to the MSG phase,
the linear form holds only for very small $\omega$, and the dynamic susceptibility
is described by the crossover function
\begin{equation}
\chi^{\prime\prime}_L ( \omega ) \sim  \frac{\omega}{\sqrt{\Delta + \sqrt{\omega^2 
+ \Delta^2}}}.
\label{cross1}
\end{equation}
The energy $ \Delta \sim r$ (recall $r$ is the tuning  parameter to the spin
glass transition) is the crossover scale separating Fermi liquid behavior
for $\omega \ll \Delta$, where (\ref{chifl}) holds, to  non-Fermi liquid behavior
for $\omega \gg \Delta$ where $\chi^{\prime\prime}_L
( \omega )
\sim \mbox{sgn} ( \omega ) \sqrt{\omega}$. 
This non-Fermi liquid behavior holds at all frequencies at the $\Delta = 0 $ critical
point; in imaginary time the critical spin correlator behaves as
\begin{equation}
\overline{\langle S_i (0) \cdot S_i (\tau) \rangle} \sim 
\frac{1}{\tau^{3/2}}. 
\end{equation}

For $r< 0$, the system is in the MSG phase. We now have
\begin{equation}
\overline{\langle S_i (0) \cdot S_i (\tau) \rangle} \sim q_{EA} +
\frac{c}{\tau^{3/2}}
\label{sg1}
\end{equation}
where $c$ is some constant, and the Edwards Anderson order parameter 
vanishes as $q_{EA} \sim r$, as one approaches the critical point to the MQP
phase. The entire structure of the Parisi spin glass order parameter~\cite{hertz}
has also been determined~\cite{rsy,sro}: the results closely parallel those of the
classical system~\cite{hertz} with replica symmetry breaking effects vanishing as
$T \rightarrow 0$ everywhere in the spin glass phase. 

Finally, it is useful to compare the $1/\tau^2$ decay of correlations in the MQP
phase with the $1/\tau$ decay found in Section~\ref{heisenberg} for the IQP phase.
We may interpret this faster decay as a consequence of the reduced local degeneracy
of each site. Each spin is coupled to a metallic fermionic bath, which,
loosely speaking,
lifts the double degeneracy on each site at asymptotically low 
energies~\cite{rmp}.

\subsection{Paired spins in an insulator\label{pair}}
This is a rather artificial model in the present context, and we discuss it only
because it is one of the simplest of the infinite range models, and closely
related models were the first
to be  solved exactly~\cite{huse,ysr,rsy}. Insights gained from their solution were
then easily transferred~\cite{sro,sg} to the metallic case already described.

Consider a random Heisenberg magnet with a natural pairing between the spins
described by the Hamiltonian $H_p$:
\begin{equation}
H_p = \sum_{i} J_{pi} S_{i1} \cdot S_{i2}
+ \sum_{i<j} \sum_{a,b=1}^{2} J_{ijab} S_{ia} \cdot S_{jb}.
\end{equation}
On each site, $i$, we now have a pair of spins $S_{i1}$ and $S_{i2}$,
and this pair interacts with an antiferromagnetic exchange $J_{pi} > 0$. 
The $J_{pi}$ are independent random variables, but they are
constrained to be positive, and have a mean value of order unity.
The couplings between spins on different sites is however similar to that in
$H$: the $J_{ijab}$ are independent Gaussian random variables with zero 
mean, and a root mean square value of order $1/\sqrt{N_s}$. 

The key property of this model is in the nature of the local spectrum at each site,
without including the consequences of the $J_{ijab}$. The ground state is a
non-degenerate singlet and it is separated by a gap, $J_{pi}$, from the 
excited
triplet state. The local degeneracy referred to in earlier subsections, 
has now
been completely lifted. It is this feature which makes the solution of this model
relatively straightforward. A similar non-degenerate local ground state, and a gap to
the lowest excited state, is
also found in spin glass models of quantum rotors~\cite{ysr} and Ising spins in a
transverse field~\cite{huse}; the properties of these models are essentially
identical to those described here.

The local spin correlations in the IQP phase now decay exponentially fast
in the infinite range model
\begin{equation}
\overline{\langle S_i (0) \cdot S_i (\tau) \rangle} \sim e^{-\Delta \tau}
\end{equation}
where $\Delta$ is an energy scale which vanishes at the transition to the ISG phase. 
The exponential decay is actually an artifact of the infinite range model,
and Griffiths effects in finite range models 
lead to a stretched exponential
contribution
$\sim \exp(- c' \sqrt{\tau})$, which dominates at long enough times.
For the local dynamic susceptibility, these statements mean that there is an
energy gap $\Delta$ in the IQP phase of the infinite-range model; 
Griffiths effects induce a weak sub-gap
absorption of order $\exp(- c'' / |\omega|)$ in models with finite range
interactions. The gap of the infinite range model $\Delta \sim  r / \ln (1/r)$ as one
approaches the critical point at $r=0$. 

Notice the faster decay of correlations in IQP phase of the present model,
as compared to those discussed earlier. This is a consequence of the 
complete lifting of the on-site degeneracy. 

Near the critical point, the crossover function
analogous to (\ref{cross1}) is
\begin{equation}
\chi^{\prime\prime}_L ( \omega ) \sim \mbox{sgn}(\omega) \sqrt{ \omega^2 - \Delta^2}
\theta(|\omega| - \Delta)
\end{equation}
So, at the critical point $\Delta = 0$, we have $\chi_L^{\prime\prime} (\omega )
\sim \omega$, and spin correlations decay as $1/\tau^2$.
In the spin glass phase, the analog of (\ref{sg1}) is
\begin{equation}
\overline{\langle S_i (0) \cdot S_i (\tau) \rangle} \sim q_{EA} +
\frac{c}{\tau^{2}};
\label{sg2}
\end{equation}
other properties of the spin glass are similar to those described in
Section~\ref{metal}.

\section{Order parameter and Landau theory}
This section reviews the results of Refs~\cite{rsy} and~\cite{sro}.

For the cases of the paired spin model of Section~\ref{pair}, and the spins
in a metal model of Section~\ref{metal}, it is possible to go beyond the
infinite-range model and study the corresponding quantum transitions 
in systems with short range interactions.  The basic strategy is similar to that
followed in the classical spin glass case: introduce an order parameter characterizing
the important long-time spin correlations, and use insights from the solution of the
infinite range model, and general symmetry considerations,
to obtain Landau functionals describing fluctuations in the 
case with finite-range interactions. 
This procedure was carried out in Ref~\cite{rsy} for the quantum rotor model
(whose properties are essentially identical to the paired spin 
model of Section~\ref{pair}), and in Ref~\cite{sro} for the metallic spin 
glass case. No such extension exists yet for the insulating Heisenberg
spin case (Section~\ref{heisenberg}), and its development remains 
an important open
problem.

In the following subsections
we will ({\em i\/}) introduce the order parameter for quantum spin glass 
to paramagnet transition~\cite{rsy}, ({\em ii\/}) obtain Landau functionals for 
finite-range versions of the models of Sections~\ref{pair}~\cite{rsy} 
and~\ref{metal}~\cite{sro},
({\em iii\/}) show
that a simple minimization of these functionals reproduces the
properties of the infinite-range models, and 
describe crossovers for a number of observables. The consequences of 
fluctuations in finite-range models is discussed in Section~\ref{scaling}.

\subsection{Order parameter} 
We begin by introducing the order parameter for the quantum phase transition~\cite{rsy}.
Recall that for classical spin glasses in the replica formalism, this is a matrix $q^{ab}$,
$a,b =1 \ldots n$ are replica indices and $n \rightarrow 0$.
The \underline{off-diagonal} components of $q^{ab}$ can be related to the
Edwards-Anderson order parameter, $q_{EA}$, in a somewhat subtle way we won't go into
here~\cite{hertz,youngbinder}. In quantum ($T=0$) phase transitions, time dependent
fluctuations of the order parameter must be considered (in ``imaginary''
Matsubara time $\tau $), and in the spin glass case it is found that the 
standard
decoupling, analogous to the classical case introducing $q^{ab}$, leads 
now to a matrix function of two times~\cite{BrayMoore} 
which we can consider to be
\begin{equation}
Q^{ab} ( x , \tau_1 , \tau_2 ) =  \sum_{i \in {\cal N}(x)}
S_{i}^a ( \tau_1 ) 
S_{i}^b (\tau_2 )
\label{composite}
\end{equation}
where ${\cal N}(x)$ is a coarse-graining region in the neighborhood of $x$.
From the set-up of the replica formalism it is clear that
\begin{eqnarray}
\overline{\langle S_i (0) \cdot S_i ( \tau ) \rangle} = 
\lim_{n \rightarrow 0}
\frac{1}{n} \sum_{a} \left\langle\left\langle Q^{aa} ( x, \tau_1=0 ,
\tau_2=\tau) \right\rangle\right\rangle  \\
q_{EA} = \lim_{\tau \rightarrow  \infty} \lim_{n \rightarrow 0}
\frac{1}{n} \sum_{a} \left\langle\left\langle Q^{aa} ( x, \tau_1=0 ,
\tau_2=\tau) \right\rangle\right\rangle 
\label{defqea}
\end{eqnarray}
relating $q_{EA}$ to the replica \underline{diagonal} components of $Q$. 
We have introduced above double angular brackets to represent averages taken with
the translationally invariant replica action (recall that single angular brackets
represent thermal/quantum averages for a fixed realization of randomness,
and overlines represent averages over randomness).
Notice that the fluctuating field $Q$ is in general a function of two separate times
$\tau_1$ and $\tau_2$; however the expectation value of its replica diagonal components
can only be a function of the time difference $\tau_1 - \tau_2$. Further, 
the expectation value of the replica
off-diagonal components of $Q$ is independent of both $\tau_1$ and $\tau_2$~\cite{ysr},
and has a structure very similar to that of the classical order parameter $q^{ab}$.
One can therefore also obtain $q_{EA}$ from the replica off-diagonal components of $Q$,
as noted above for $q^{ab}$~\cite{hertz,youngbinder}. Let us also note for completeness
that, unlike the quantum case, the replica diagonal components of the 
classical order parameter $q^{ab}$ are 
usually constrained to be unity~\cite{hertz}, and contain no useful information.

The order parameter we shall use is $Q^{ab} (x, \tau_1, \tau_2)$, which is
a matrix in a replica space
and depends on the spatial co-ordinate $x$ and two times $\tau_1$, $\tau_2$. However,
a little thought shows that this function contains too much information. The important
degrees of freedom, for which one can hope to make general and universal statements,
are the long-time spin correlations with $|\tau_1 - \tau_2| \gg \tau_m$,
where $\tau_m$ is a microscopic time like an inverse of a typical 
exchange constant. As presented, the function $Q$ contains
information not only on the interesting long-time correlations,
but also on the uninteresting time range with $|\tau_1 - \tau_2|$
smaller than or of order $\tau_m$. The correlations in the latter range
are surely model-dependent and cannot be part of any general Landau action.
We shall separate out this uninteresting part of $Q$ by performing the shift
\begin{equation}
Q^{ab} (x, \tau_1 , \tau_2 ) \rightarrow
Q^{ab} (x, \tau_1 , \tau_2 ) - C \delta^{ab} \delta (\tau_1 - \tau_2)
\label{repar}
\end{equation}
where $C$ is a constant, and the delta function $\delta ( \tau_1 - \tau_2 )$
is a schematic for a function which decays rapidly to zero on a scale $\tau_m$.
The value of $C$ will be adjusted so that the resulting $Q$ contains only
the interesting long time physics: we will see later how this can be done
in a relatively straightforward manner. 
The alert reader may recognize some similarity between the above procedure,
and M.E. Fisher's analysis~\cite{mef} of the Yang-Lee edge problem.
In that case, too, the order parameter contains an uninteresting non-critical
piece which has to be shifted away; we will see below that there many other
similarities between the Yang-Lee edge and quantum spin glass problems.

\subsection{Action functional\label{action}}
The action functional can be derived by explicit computations on  
microscopic models or deduced directly from general arguments which
have been discussed in some detail in Ref~\cite{rsy}. 
Apart from a single non-local term present in the metallic case~\cite{sro}
(see below), the remaining important terms are consistent with the general
criteria~\cite{rsy} that:
\begin{enumerate}
\item The action is an integral over space of a local operator which can be expanded in
gradients of powers of $Q$ evaluated at the same position $x$.
\item $Q$ is bilocal ({\em i.e.}\ is a matrix) in time, and each time is associated with
one of the two replica indices (see
definition Eqn (\ref{composite})). These ``indices'' can appear more than once in a term and
are summed over freely subject to the following rules before summations:
\begin{enumerate}
\item Each distinct replica index appears an even number of times~\cite{hertz}.
\item Repetition of a time ``index'' corresponds to quantum-mechanical
interaction of spins, which must be
local in time and accordingly can be expanded as terms with times set 
equal plus the same with
additional derivatives; it occurs when the corresponding replica 
indices are the same, and only
then.
\end{enumerate}
\end{enumerate}

We now present all the terms, which, a subsequent renormalization-group 
analysis tells
us are important near the quantum critical point. This is only a small subset of
the terms allowed by the above criteria.

A crucial term is that linear in the order parameter $Q$. This term
encodes the local, on-site physics of the spin glass model. To the order
we shall consider, this is the only term which distinguishes between the
finite-range analogs of the models of Section~\ref{metal} and~\ref{pair}.
For the paired spin model of Section~\ref{pair} 
we find the linear term~\cite{rsy}
\begin{equation}
\frac{1}{\kappa t} \int d^d x  d\tau \sum_a
\left. \left[ \frac{\partial}{\partial\tau_1}
\frac{\partial}{\partial \tau_2} + \widetilde{r} \right] Q^{aa} (x , \tau_1 , \tau_2 )
\right|_{\tau_1=\tau_2=\tau} 
\label{actione1}
\end{equation}
The gap in the on-site spectrum for this model 
tells us that it is valid to expand the linear term in derivatives of $\tau$,
and the above contains the leading order contributions. The coupling
$\widetilde{r}$ will be seen below to be the critical tuning parameter for the transition
from the spin glass to the paramagnet (it is related to the parameter $r$,
introduced earlier, by an additive constant). 
There is an overall factor of $1/\kappa t$
in front of this term; we have written this factor as a product of two coupling
constants, $\kappa$ and $t$, for technical reasons~\cite{rsy} we shall not discuss
here. Turning to the metallic model (Section~\ref{metal}) we find, instead,
the linear term~\cite{sro}
\begin{equation}
\frac{1}{\kappa t} \int d^d x \left\{ \int d\tau
\sum_a r Q^{aa} (x , \tau , \tau ) - \frac{1}{\pi} \int
d\tau_1 d\tau_2
\sum_{a} \frac{Q^{aa} (x, \tau_1, \tau_2)}{(\tau_1 - \tau_2 )^2}
\right\}
\label{actione2}
\end{equation}
Now the coupling of the local spin degree of freedom to the itinerant quasiparticles
has introduced a long range $1/\tau^2$ interaction in imaginary time: this is clearly
related to the result (\ref{metale1}). A term with two time derivatives, like
that in (\ref{actione1}), is also permitted, but it is not as important at
long times. 

The remaining terms presented here
apply to both the insulating paired spins and spins in a metal
model.

There is a quadratic gradient term
\begin{equation}
\frac{1}{2t} \int  d^d x d \tau_1
d \tau_2 \sum_{a,b} \left[ \nabla Q^{ab} (x, \tau_1, \tau_2 )
\right]^2
\label{actione3} 
\end{equation}
which is responsible for the development of spatial correlations in the 
spin glass order. Clearly, such a term is absent in the infinite-range case,
and is special to the more realistic short-range case. A quadratic term
without gradients
\begin{equation}
\int d^d x d \tau_1
d \tau_2 \sum_{a,b} \left[ Q_{\mu\nu}^{ab} (x, \tau_1, \tau_2 )
\right]^2
\label{actione4}
\end{equation}
is also allowed by the general criteria, but we choose to tune its coefficient to
zero by using the freedom in (\ref{repar}). As will become clear in the next
subsection, this criterion is identical to requiring the absence of
uninteresting short time behavior in $Q$.
Notice again the formal similarity to the theory of the Yang-Lee edge~\cite{mef},
where setting the coefficient of a quadratic term to zero was also responsible
for removing the uninteresting non-critical part of the order parameter variable.

Next we consider cubic non-linearities, and the most important among the several
allowed terms is the one with the maximum
number of different time and replica indices:
\begin{equation}
- \frac{\kappa}{3t} \int  d^d x d \tau_1 d \tau_2 d \tau_3
\sum_{a,b,c} Q^{ab} (x, \tau_1 , \tau_2 ) Q^{bc}
(x, \tau_2 , \tau_3 ) Q^{ca}
(x, \tau_3 , \tau_1 ). 
\label{actione5}
\end{equation}
This term accounts for non-linearities induced by solely by disorder 
fluctuations.

Of the terms with fewer than the maximum allowed number of time indices
at a given order, the most important one is the one at quadratic order:
\begin{equation}
\frac{u}{2t} \int d^d x d \tau \sum_a  u~Q^{aa} ( x, \tau , \tau) Q^{aa} 
( x, \tau , \tau).
\label{actione6}
\end{equation}
We have ignored vector spin indices ($\mu,\nu$) here, and if these were 
accounted, we would have found two separate terms with the same
replica and time integration structure as (\ref{actione6})~\cite{rsy}; 
however the
additional term does not significantly modify the physics, and we
will ignore it here. The coupling $u$ is the only one responsible for quantum mechanical
interactions between the spins, and as a consequence, all the time and 
replica indices in (\ref{actione6}) are the same.

Finally we have a final quadratic term
\begin{equation}
- \frac{1}{2t^2} \int d^d x \int  d \tau_1 d \tau_2
\sum_{a,b}
Q^{aa}  (x, \tau_1 , \tau_1 ) Q^{bb} ( x, \tau_2 , \tau_2 ),
\label{actione7}
\end{equation}
which accounts for the spatial fluctuation in the position of the paramagnet-spin glass
transition. Recall that the linear coupling $\widetilde{r}$ was the control parameter for this
transition, and a term like (\ref{actione7}) is obtained by allowing for Gaussian 
fluctuations in $\widetilde{r}$, 
about its mean value, from point to point in space.
It will turn out that (\ref{actione7}) plays no role in the mean-field analysis
in the following subsection. However, it is essential to include (\ref{actione7})
for a proper theory of the fluctuations, and its effects are responsible for
ensuring that the correlation length exponent, $\nu$, satisfies the inequality~\cite{chayes}
$\nu > 2/d$. 

To summarize this subsection, the action functional for the paired spin model
with short-range interactions is (\ref{actione1}) + (\ref{actione3}) + (\ref{actione5}) 
+ (\ref{actione6})
 + (\ref{actione7}), while that for the spins in a metal case is
(\ref{actione2}) + (\ref{actione3}) + (\ref{actione5}) 
+ (\ref{actione6})
 + (\ref{actione7}).
Notice that the two differ only in the form of the linear term, which is
(\ref{actione1}) for the paired spin case, and (\ref{actione2}) for the metallic
case.

\subsection{Mean field theory\label{mean}}

We will limit our discussion in this subsection to the metallic case. We will review
the mean field theory for the MQP phase~\cite{sro,sg} and identify the position of its
instability to the MSG phase. A discussion of the solution within the MSG phase
will not be presented here, and appears elsewhere~\cite{sro}.

We Fourier transform from imaginary time to Matsubara frequencies by expressing
the action in terms of
\begin{equation}
Q^{ab} (x, \omega_1 , \omega_2 ) = \int_0^{1/T} d \tau_1 d \tau_2
Q^{ab} (x, \tau_1 , \tau_2 ) e^{- i ( \omega_1 \tau_1 + \omega_2 \tau_2 )},
\label{mft1}
\end{equation}
where we are using units in which $\hbar = k_B = 1$, and the frequencies,
$\omega_1$, $\omega_2$ are quantized in integer multiples of $2 \pi T$. 
Then, we 
make an ansatz for the mean-field value of $Q$ which is $x$-independent, and
dependent only on $\tau_1 - \tau_2$; within the MQP phase this takes the form
\begin{equation}
Q^{ab} (x, \omega_1, \omega_2 ) = (\delta^{ab} \delta_{\omega_1 + 
\omega_2, 0}/T ) \chi_L (i \omega_1 )
\label{mft2}
\end{equation}
where we have used (\ref{defqea}) to identify the right hand side as the local dynamic
susceptibility. Inserting (\ref{mft2}) into the action for the metallic case
in Section~\ref{action}, we get for the free energy per unit volume 
${\cal F}/n$
(as usual~\cite{hertz}, ${\cal F}/n$ represents the physical disorder averaged free energy):
\begin{equation}
\frac{{\cal F}}{n} = \frac{T}{t} \sum_{\omega} \left[
\frac{|\omega| + \widetilde{r}}{\kappa} \chi_L (i \omega )  - \frac{\kappa}{3} \chi_L^3 (i \omega )
\right] + \frac{u}{2t} \left[ T \sum_{\omega} \chi_L ( i \omega ) \right]^2
\label{mft3}
\end{equation}
Notice that the coupling $1/t$ appears only as a prefactor in front of the total 
free energy, as the contribution of the $1/t^2$ term (\ref{actione7}) vanishes in the replica
limit $n \rightarrow 0$. The value of $t$ will therefore play no role in the mean field 
theory. We now determine the saddle point of (\ref{mft3}) with respect to variations
in the whole function $\chi_L ( i \omega )$, and find the solution
\begin{equation}
\chi_L ( i \omega ) = - \frac{1}{\kappa} \sqrt{ |\omega | + \Delta},
\label{mft4}
\end{equation}
where the energy scale $\Delta$ is determined by the solution of the equation
\begin{equation}
\Delta = \widetilde{r} - u T \sum_{\omega} \sqrt{|\omega | + \Delta }.
\label{mft5}
\end{equation}
Taking the imaginary part of the analytic continuation of (\ref{mft4})
to real frequencies, we get 
\begin{equation}
\chi_L^{\prime\prime} ( \omega ) = \frac{1}{\kappa \sqrt{2}}
\frac{\omega}{\sqrt{\Delta + \sqrt{\omega^2 + \Delta^2}}},
\label{cross2}
\end{equation}
which was also the result (\ref{cross1})
for the infinite range model.
Inserting the solution for $\chi_L$ back into (\ref{mft3}), and using 
(\ref{mft5}), we get for the 
free energy density:
\begin{equation}
\frac{{\cal F}}{n} = - \frac{1}{\kappa^2 t} \left[
\frac{2T}{3} \sum_{\omega} ( |\omega| + \Delta )^{3/2} + \frac{u}{2}
\left( T \sum_{\omega} \sqrt{ |\omega | + \Delta} \right)^2 \right] 
\label{mft5a}
\end{equation}
The equations (\ref{mft5}), (\ref{cross2}), (\ref{mft5a}) are key 
results~\cite{sro,sg}, from which our mean field predictions for physical observables
will follow. 
Despite their apparent simplicity, these results contain a great
deal of structure, and a fairly careful and non-trivial analysis is required
to extract the universal information contained within them.

First, it is easy to note that there is no sensible solution (with $\Delta > 0$)
of (\ref{mft5}) at $T=0$ for $\widetilde{r} < \widetilde{r}_c$ where
\begin{equation}
\widetilde{r}_c = u \int \frac{d \omega}{2 \pi} \sqrt{|\omega|} \approx \frac{2 \Lambda_{\omega}^{3/2}}{
3 \pi}
\label{mft6}
\end{equation}
where $\Lambda_{\omega}$ is an upper cut-off in frequency. Clearly the system is in the
MSG phase for $T=0$, $\widetilde{r} < \widetilde{r}_c$, and a separate ansatz for $Q$ is
necessary  there, as discussed
elsewhere~\cite{sro}. Let us now define
\begin{equation}
r \equiv \widetilde{r} - \widetilde{r}_c,
\label{mft7}
\end{equation}
so that the quantum critical point is at $T=0$, $r=0$. In the vicinity of 
this point, our action constitutes a continuum quantum field theory (CQFT)
describing the physics of the system at all energy scales significantly
smaller than $\Lambda_{\omega}$. The ``universal'' properties of the system
are the correlators of this CQFT, and they apply therefore for 
$r, T \ll \Lambda_{\omega}$, a condition we assume in our analysis
below. It is also natural to assume that the microscopic coupling
$u \sim \Lambda_{\omega}^{-1/2}$. We shall, however, make no assumptions
on the relative magnitudes of $r$ and $T$.

Let us now examine the solution of (\ref{mft5}) under the conditions
noted above.
Clearly, we have from (\ref{mft5}) and (\ref{mft7}) that
\begin{equation}
\Delta + u T \sqrt{\Delta} = 
r - u \left( T \sum_{\omega \neq 0 } \sqrt{|\omega | + \Delta }
- \int \frac{d \omega}{2 \pi} \sqrt{|\omega|} \right).
\label{mft8}
\end{equation}
We have chosen to move the $\omega = 0$ term in the
frequency summation from the right hand to the left hand side.
This permits us to replace the $\Delta$ on the right hand side by its
zeroth order result in an expansion in $u$, as all neglected terms can be
shown to be less singular as one approaches the critical point. After doing this,
we further manipulate (\ref{mft8}) into
\begin{eqnarray}
\Delta + u T \sqrt{\Delta} = 
r + u T \sqrt{r} &-& u \left(T \sum_{\omega} \sqrt{|\omega | + r }
- \int \frac{d \omega}{2 \pi} \sqrt{|\omega| + r} \right) \nonumber \\
&-& u \int \frac{d \omega}{2 \pi}\left( \sqrt{|\omega | + r }
- \sqrt{|\omega|} - \frac{r}{2 \sqrt{|
\omega|}} \right) \nonumber \\
&-& u \int \frac{r}{2 \sqrt{|\omega|}}.
\label{mft9}
\end{eqnarray}
In the first term on the right hand side, the objective has been to always
subtract from the  summation over Matsubara frequencies
of any function, the integration of precisely the same function; the 
difference is then strongly convergent in the ultraviolet, and 
such a procedure leads
naturally to finite temperature crossover functions~\cite{epsilon}. The second term
also has a subtraction to make it convergent in the ultraviolet.
We will now manipulate (\ref{mft9}) into a form  where it is evident
that $\Delta$ is analytic as a function of $r$ at $r=0$ for $T \neq 0$.
There is a thermodynamic singularity at the quantum critical point
$T=0$, $r=0$, but this must disappear at $r=0$ for any non-zero $T$~\cite{epsilon}
(see Fig~\ref{f2}).
We use the identity
\begin{equation}
\int_0^{\infty} \sqrt{s} ds \left( \frac{1}{s + a} - \frac{1}{s + b} \right)
= \pi ( \sqrt{b} - \sqrt{a})
\end{equation}
to rewrite (\ref{mft9}) as 
\begin{eqnarray}
\Delta + && u T \sqrt{\Delta} =  r\left( 1 - \frac{u \Lambda_{\omega}^{1/2}}{\pi}
\right)  + u T \sqrt{r} \nonumber \\
&&~~~ + \frac{u}{\pi} \int_0^{\infty} \sqrt{s} ds \left(
T \sum_{\omega} \frac{1}{s + |\omega| + r} - \int \frac{d\omega}{2 \pi}
\frac{1}{s + |\omega| + r} \right) \nonumber \\
&&~~~ + \frac{u}{\pi} \int_0^{\infty} \sqrt{s} ds \int \frac{d\omega}{2 \pi}
\left(
\frac{1}{s + |\omega| + r} - \frac{1}{s + |\omega| }
+ \frac{r}{(s + |\omega|)^2}\right).
\end{eqnarray}
We now evaluate the frequency summation by expressing it in terms of the
digamma function $\psi$, and perform all frequency integrals exactly.
After some elementary manipulations (including use of the identity
$\psi(s+1 ) = \psi(s) + 1/s$), we obtain our final result for $\Delta$, in the form
of a solvable quadratic equation for $\sqrt{\Delta}$:
\begin{equation}
\Delta + u T \sqrt{\Delta} = 
r \left( 1 - \frac{u \Lambda_{\omega}^{1/2}}{\pi} \right)
+ u T^{3/2} \Phi \left(\frac{r}{T} \right),
\label{mft9a}
\end{equation}
where the universal crossover function $\Phi(y)$ is given by
\begin{equation}
\Phi (y) = 
\frac{1}{\pi^2} \int_0^{\infty}
\sqrt{s} ds \left[ 
\log\left( \frac{s}{2 \pi} \right) - \psi \left( 1 + \frac{s+y}{2 \pi} \right)
+ \frac{\pi + y}{s} \right].
\label{mft10}
\end{equation}
The above expression for $\Phi(y)$ is clearly analytic for all $y \geq 0$, including
$y=0$, as required from general thermodynamic considerations~\cite{epsilon}.
Indeed, we can use the above result even for $y<0$ until we hit the
first singularity  at $y=- 2\pi$, which is associated with singularity
of the digamma function $\psi(s)$ at $s=0$. However, this singularity is of no
physical consequence, as it occurs within the spin glass phase (F ig~\ref{f2}), where
the above solution is not valid; as shown below, the transition to the spin
glass phase occurs for $y \sim - u T^{1/2}$ which is well above $-2 \pi$.
It is useful to have the following limiting results,
which follow from (\ref{mft10}), for our subsequent analysis:
\begin{equation}
\Phi(y) = \left\{
\begin{array}{cc}
\sqrt{1/2\pi} \zeta(3/2) + {\cal O}(y) & y \rightarrow 0 \\
(2 /3 \pi) y^{3/2} + y^{1/2} + (\pi/6) y^{-1/2}  + {\cal O}(y^{-3/2})
& y \rightarrow \infty 
\end{array}
\right.
\end{equation}

The expression (\ref{cross2}),
combined with the results (\ref{mft9a}) and (\ref{mft10}) 
completely specify the $r$ and $T$ dependence of the dynamic susceptibility
in the MQP phase, and allow us to obtain the phase diagram shown in Fig~\ref{f2}.
The crossovers shown are properties of the CQFT
characterizing the quantum critical point. We present below explicit results
for the crossover functions of a number of observables 
within the mean field theory. 
A more general scaling interpretation
will be given in Section~\ref{scaling}. 

Before describing the crossovers, we note that
the full line in Fig.~\ref{f2} denotes the
boundary of the paramagnetic phase at $r = r_c (T)$ (or $T = T_c (r)$).
This is the only line of thermodynamic phase transitions, and its location is
determined by the condition $\Delta = 0$, which gives us 
\begin{equation}
r_c (T) =  - u \Phi(0) T^{3/2}~~~\mbox{or}~~~T_c (r) = (-r / u \Phi (0))^{2/3}
\label{mft10a}
\end{equation}

The different regimes in Fig~\ref{f2} can be divided into two classes determined
by whether $T$ is ``low'' or ``high''. There are two low $T$ regimes, one for
$r>0$, and the other for $r<0$; these regions display properties of the
non-critical ground states, which were reviewed in Section~\ref{general}. 
More novel is the high $T$ 
region, where the most important energy scale is set by $T$, and 
``non-Fermi liquid'' effects associated with the critical ground state 
occur. We now describes the regimes in more detail, in turn.
\newline
(I) \underline{Low $T$ region above MQP ground state, $T < (r/u)^{2/3}$}
\newline
This is the ``Fermi liquid'' region, where the leading contribution
to $\Delta$ is its $T=0$ value $\Delta(T) \sim \Delta (0) = r$.
The leading temperature dependent correction to $\Delta$ is
however different in two subregions.
In the lowest $T$ region Ia, $T < r$, we have the Fermi liquid $T^2$
power law
\begin{equation}
\Delta (T) - \Delta (0) = \frac{u \pi T^2}{6 \sqrt{r}}~~~~~~~~\mbox{region Ia}.
\end{equation}
At higher temperatures, in region Ib, $r < T < (r/u)^{2/3}$,
we have an anomalous temperature dependence
\begin{equation}
\Delta (T) - \Delta (0) = u \Phi (0) T^{3/2}~~~~~~~~\mbox{region Ib and II}.
\label{mft11}
\end{equation}
It is also interesting to consider the properties of  region I as a function
of observation frequency, $\omega$, as sketched in Fig.~\ref{f3}.
At large frequencies, $\omega \gg r$, the local dynamic susceptibility
behaves like $\chi_L^{\prime\prime} \sim \mbox{sgn}(\omega) \sqrt{|\omega|}$,
which is the spectrum of critical fluctuations; at the $T=0$, $r=0$ critical point,
this spectrum is present at all frequencies. At low frequencies, $\omega \ll r$,
there is a crossover (Fig~\ref{f3}) to the characteristic Fermi liquid spectrum
of local spin fluctuations $\chi_L^{\prime\prime} \sim \omega/\sqrt{r}$.
\newline
(II) \underline{High $T$ region, $T > (|r|/u)^{2/3}$}
\newline   
This is the ``non-Fermi liquid'' region, where temperature dependent
contributions to $\Delta$ dominate over those due to the deviation
of the coupling $r$ from its critical point, $r=0$. Therefore thermal effects
are dominant, and the system behaves as if its microscopic couplings are
at those of the critical ground state. The $T$ dependence in 
(\ref{mft11}) continues
to hold, as we have already noted, with the leading contribution now being
$\Delta \approx u \Phi (0) T^{3/2}$.
As in (I), it is useful to consider properties of this region as a 
function of $\omega$ (Fig~\ref{f3}).
For large $\omega$ ($ \omega \gg u T^{3/2}$) we again have the critical behavior
$\chi_L^{\prime\prime} \sim \mbox{sgn} (\omega ) \sqrt{|\omega|}$; this critical behavior
is present at large enough $\omega$ in all the regions of the phase diagram.
At small $\omega$ ($ \omega \ll u T^{3/2}$), thermal fluctuations quench the 
critical fluctuations, and we have relaxational behavior with
$\chi_L^{\prime\prime} \sim \omega / u^{1/2} T^{3/4}$.
\newline
(III) \underline{Low $T$ region above MSG ground state, $T < (-r/u)^{2/3}$}
\newline
Effects due to the formation of a static moment are now paramount.
As one approaches the spin glass boundary (\ref{mft10a}) from above,
the system enters a region of purely classical thermal fluctuations,
$|T - T_c (r) | \ll u^{2/3} T_c^{4/3} (r)$
(shown shaded in Fig~\ref{f2}) where
\begin{equation}
\Delta = \left(\frac{r-r_c (T)}{T u} \right)^2
\end{equation}
Notice that $\Delta$ depends on the square of the distance from the
finite $T$ classical phase transition line, in contrast to its linear
dependence, along $T=0$, on the deviation from the quantum critical point 
at $r=0$.

We have now completed a presentation of the mean field predictions for the finite $T$
crossovers near the quantum critical point (Fig~\ref{f2}), and for the explicit crossover
functions for the frequency-dependent local dynamic susceptibility (Fig~\ref{f3}
and Eqns (\ref{cross2}), (\ref{mft9a}), (\ref{mft10})).
We will now consider implications of our results for a number of 
other experimental observables.

\subsubsection{Nuclear relaxation:}
This was considered in Ref~\cite{sg}.
The $1/T_1$ relaxation rate of nuclei coupled to the electronic spins by a
hyperfine coupling is given by the low frequency limit of the local 
dynamic susceptibility.
We have
\begin{equation}
\frac{1}{T_1 T} = A^2 \lim_{\omega \rightarrow 0} \frac{\chi_L^{\prime\prime} (\omega)}{
\omega} = \frac{A^2}{2 \kappa \sqrt{\Delta}},
\end{equation}
where $A$ is determined by the hyperfine coupling. The crossover function
for $1/T_1$ now follows from that for $\Delta$ in (\ref{mft9a}), (\ref{mft10}).
In particular, in region II, $1/T_1 \sim T^{1/4}$~\cite{sg}.
 
\subsubsection{Uniform linear susceptibility:}
In the presence of a uniform magnetic field, $H$, the action acquires the term~\cite{sro}
\begin{equation}
- \frac{g}{2t} \int d^d x d \tau_1 d \tau_2 \sum_{ab\mu\nu} Q^{ab}_{\mu \nu}
(x, \tau_1 , \tau_2 ) H_{\mu} H_{\nu},
\label{expt1}
\end{equation}
where $\mu$, $\nu$ are spin indices. There are a number of additional terms
involving $H$~\cite{sro}, but their linear response is always weaker 
than that due to (\ref{expt1}). Taking the second derivative of the 
free energy with respect to $H$, we obtain for the uniform 
linear susceptibility, $\chi_u$~\cite{rsy,sg}
\begin{equation}
\chi_u  = \chi_b - \frac{g}{t \kappa} \sqrt{\Delta},
\end{equation}
where $\chi_b$ is the $T$, $r$- independent background contribution of the fermions
that have been integrated out.
The crossover function for $\chi_u$ now follows, as for $1/T_1$, from that for $\Delta$.

\subsubsection{Nonlinear susceptibility:}
The non-linear response to (\ref{expt1}) was computed in Ref~\cite{rsy}. The nonlinear
susceptibility was found to be
\begin{equation}
\chi_{nl} = \frac{u g^2}{4t} \frac{1}{\Delta},
\end{equation}
and its crossover function again follows from that for $\Delta$. Notice that $\chi_{nl}$
is proportional to the quantum mechanical interaction $u$, and would vanish in a 
theory with terms associated only with disorder fluctuations.

\subsubsection{Free energy and specific heat:}
The result for the free energy was given in (\ref{mft5a}), and it needs to be
evaluated along the lines of the analysis carried out above for the crossover
function determining $\Delta$. The specific heat then follows by the usual thermodynamic
relation. 
After performing the necessary frequency summations and integrations,
we obtained the crossover function for the free energy density:
\begin{eqnarray}
\fl \frac{{\cal F} (T, r) - {\cal F}(T=0,r=0)}{n} = 
-\frac{1}{\kappa^2 t} \left[ \frac{2 r \Lambda_{\omega}^{3/2}}{3 \pi} \right.
 &+& 
T^{5/2} \Phi_{{\cal F}} \left( \frac{\Delta}{T} \right)
+ \frac{\Lambda_{\omega} \Delta^2}{2 \pi}  \nonumber \\
&+& \left. \frac{(\Delta - r)^2}{2 u} - \frac{4 \Delta^{5/2}}{15 \pi} \right],
\end{eqnarray}
where
\begin{equation}
\Phi_{{\cal F}} (y) = - \frac{2 \sqrt{2}}{3 \pi}
\int_0^{\infty} \frac{ d \Omega}{e^{\Omega} 
- 1} \frac{ \Omega (2 y + \sqrt{\Omega^2 + y^2})}{
\sqrt{ y + \sqrt{ \Omega^2 + y^2}}}
\end{equation}
This result for ${\cal F}$ includes non-singular contributions,
smooth in $r$, which form a background to the singular critical 
contributions. In region II, the most singular term is the one
proportional to $\Phi_{\cal F}$, and yields a specific heat, $C_v$~\cite{sg}
\begin{equation}
\frac{C_v}{T} = \gamma_b - \frac{\zeta (5/2) }{\sqrt{2 \pi} \kappa^2 t} \sqrt{T}
\end{equation}
where $\gamma_b$ is a background contribution.

\subsubsection{Charge transport:}
The consequences of the order parameter fluctuations on charge transport
were explored by Sengupta and Georges~\cite{sg}. The quasiparticles are
assumed to carry both charge and spin, and they scatter off the
spin fluctuations via an exchange coupling. In the Born approximation,
this leads to a contribution to the quasi particle relaxation rate,
$1/\tau_{qp}$
\begin{equation}
\frac{1}{\tau_{qp}} \propto \int_0^{\infty} \frac{d \Omega}{\sinh(\Omega/T)}
\frac{\Omega}{\sqrt{\Delta + \sqrt{\Omega^2 + \Delta^2}}},
\end{equation}
whose $T$ and $r$ dependence follows from that of $\Delta$.
In region II, $1/\tau_{qp} \sim T^{3/2}$.

\section{Fluctuations and scaling analysis\label{scaling}}
Fluctuations about the mean field theory just described have been
discussed in some detail in Refs~\cite{rsy,sro}, and we will be quite
brief here. 
As in Section~\ref{mean},
we will also restrict all our discussion to the metallic spin glass case.

The structure of the corrections can be understood
by examining the behavior of the action under the renormalization
group transformation
\begin{equation}
x \rightarrow x e^{-\ell}~~~~~~~\tau \rightarrow \tau e^{-z \ell}.
\end{equation}
The mean field theory has dynamic exponent $z=4$. 

The system is tuned across the transition by the control parameter $r$,
and by the definition of the correlation length exponent, $\nu$, it obeys 
\begin{equation}
\frac{d r}{d \ell} = \frac{1}{\nu} r + \ldots
\end{equation}
In mean field theory we have $\nu = 1/4$.

The quantum mechanical interaction, $u$, obeys for weak coupling
\begin{equation}
\frac{d u}{d \ell } = - \theta_u u,
\end{equation}
with $\theta_u = 2$. So the mean field theory in fact corresponds to a 
fixed point with $u^{\ast} = 0$, and all $u$ dependent corrections
are, in some sense, corrections to scaling. It is essential to include
such corrections, as none of the remaining couplings involve quantum mechanical
interactions between the order parameter modes, and therefore lead to
contributions insensitive
to the value of $T$. 

The parameter $t$ controls fluctuations of disorder, and for weak-coupling
its renormalization group equation is
\begin{equation}
\frac{dt}{d\ell} = - \theta t
\label{rgt}
\end{equation}
with $\theta = 2$. The value of $t$ was immaterial in the mean-field theory,
and it might appear from the flow (\ref{rgt}) that we can set $t=0$ in the 
analysis of the fluctuations. This not strictly correct, as many quantities
have a singular dependence on $t$ as $t \rightarrow 0$. As a result,
$t$ behaves like a ``dangerously irrelevant'' variable and modifies
``hyperscaling'' relations between exponents. These effects are discussed
in more detail in Ref~\cite{rsy,sro}, and as they are peripheral to the
emphasis here, we will not say more.

Finally, the cubic non-linearity, $\kappa$, obeys
\begin{equation}
\frac{d \kappa}{d \ell} = \frac{8-d}{2} \kappa + 9 \kappa^3
\end{equation}
For $d > 8$, there is a stable fixed point at $\kappa = 0$, which
describes the mean field solution discussed in Section~\ref{mean}.
However, for the physically interesting $d<8$, this fixed point 
is unstable and the system has a runaway flow to strong coupling.
So we have no systematic way of controlling the corrections
due to fluctuations. 

Under these circumstances, the best we can hope to do is to use
the insight gained from the mean field solution to speculate
on some reasonable scaling scenarios. Two natural possibilities
arise, depending upon whether the fixed point values of $u$ and other
quantum mechanical interactions are all zero (a {\em static \/} fixed 
point) or not (a {\em dynamic\/} fixed point).  
The static fixed point will have a $\theta_u > 0$; more generally
$\theta_u$ is the negative of the renormalization group eigenvalue
of the least irrelevant of the quantum mechanical couplings. 
Observables sensitive to the order parameter modes will have
their $T$ dependence controlled by the value of $\theta_u$.
Dynamic fixed points have non-zero quantum interactions
already in the critical theory, and their finite $T$
crossovers were reviewed recently in Ref~\cite{statphys};
they correspond to the $\theta_u = 0$ case of the finite $T$
behaviors of static fixed points, discussed below.

The remaining discussion summarizes a scaling interpretation
of the results of the mean-field theory in terms of
a static fixed point; we propose that behavior of a generic static
fixed point will be similar, but with different numerical values for
the exponents. The crossovers are interpreted in terms
of the exponents $z$, $\nu$, $\theta$, $\theta_u$ and an order parameter
anomalous dimension $\eta$. Recall that the mean field theory
has $z=4$, $\nu=1/4$, $\theta = 2$, $\theta_u = 2$, and $\eta = 0$.
Using arguments presented in Ref~\cite{sro}, we can deduce the 
expected behavior of all the phase boundaries in Fig~\ref{f2}
at the inaccessible (but presumed static) fixed point, in terms of the
above exponents; the results
are summarized in the caption to Fig~\ref{f2}. 
Within region II of Fig~\ref{f2}, we obtain the following $T$ dependencies
for the observables considered in the mean field theory:
\begin{eqnarray}
\frac{1}{T_1 T} \sim T^{(d-\theta-2-2z+\eta)(1 + \theta_u \nu)/2 z}
\nonumber \\
\chi_u \sim T^{(d-\theta-2+\eta)(1 + \theta_u \nu)/2 z} \nonumber \\
\chi_{nl} \sim T^{-(2 + z - \eta - \theta_u)(1 + \theta_u \nu)/z}
\nonumber \\
\frac{C_v}{T} \sim T^{(d-z-\theta)/z} \nonumber \\
\frac{1}{\tau_{qp}}  \sim  T^{(d-\theta +2 z - 2 + \eta)/2z}
\end{eqnarray}
The mean field results are obtained by using the mean field exponents
and setting $d=8$.

Ref~\cite{sro} presented additional speculations on the static fixed point
by examining the structure of the perturbation theory of the action
of Section~\ref{action}. It was found that the most singular terms
in the perturbation series were identical to those of a much simpler {\em 
classical} statistical mechanics problem: that of the Yang-Lee edge
in a random Ising ferromagnet. This mapping, and some additional assumptions,
led to the exponent relations
\begin{eqnarray}
z \nu &=& 1 \nonumber \\
\frac{1}{\nu} &=& \frac{d-\theta + 2 - \eta}{2}
\end{eqnarray}
As expected, these relations are satisfied by the mean field theory in 
the upper critical dimension $d=8$.

\section{Conclusions}
It should be clear that much additional work remains to be done,
especially with regard to understanding the consequences of disorder-induced
fluctuations. Among the issues that could be looked at in future
work are:
\begin{itemize}
\item
Is the quantum critical point for the MSG to MQP transition purely static,
{\em i.e.} does it have no quantum mechanical interactions between the 
order parameter modes ?
\item
A particular feature of our mappings to static critical points is that
all exponents and crossovers are insensitive to the number of spin
components of the order parameter, $M$, at least at all fixed points
accessible within perturbation theory. The mapping to the random Yang-Lee 
edge, if correct, would also imply an independence on $M$.
More recently, Senthil and Majumdar~\cite{senthil}, have 
exactly obtained the critical properties of 
a number of random, insulating, quantum spin chains, and found that the  
critical exponents are also independent of $M$. An important open question, 
then, is whether this independence also holds for realistic metallic and 
insulating quantum spin glasses.
\item
It would be useful to study the classical, random, Yang-Lee edge problem in 
more detail. In particular, although this is a non-trivial problem in $d=2$,
the critical theory is expected to be conformally invariant, and progress
may be possible using modern methods for such systems.
\item
While we have a complete mean field theory for the MSG to MQP transition,
there is no such theory for the ISG to IQP transition. The 
IQP to ISG transition is not understood even in the infinite-range
model, and further progress should be possible. 
\end{itemize}

\ack
We are grateful to our collaborators
on the work reviewed here, J. Ye and R. Oppermann. We thank J. Mydosh and
T. Senthil for
valuable comments on the manuscript.
This research was supported by the U.S. National Science Foundation
under grant number DMR-96-23181.

\begin{figure}
\epsfxsize=5.5in
\centerline{\epsffile{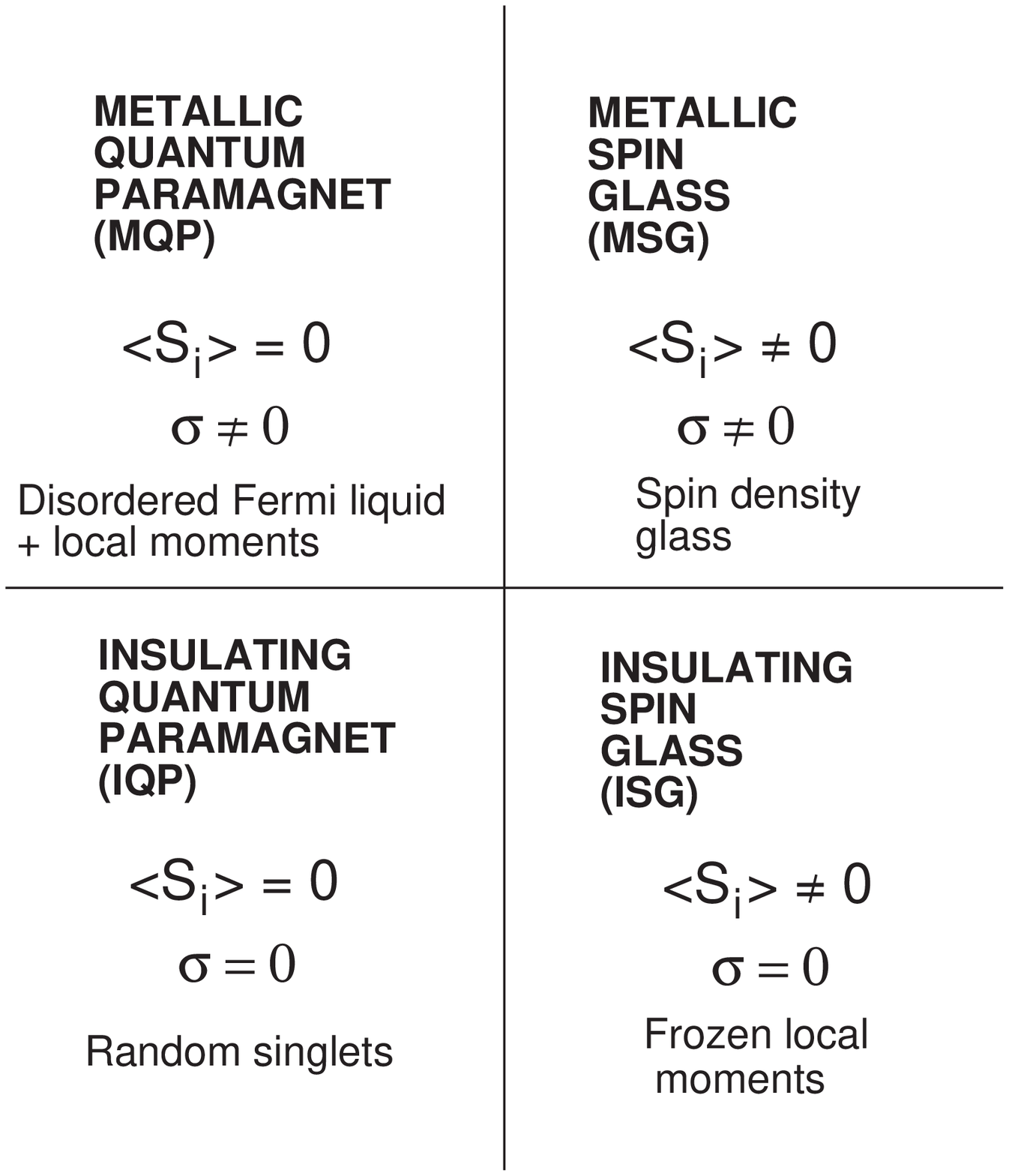}}
\vspace{0.5in}
\caption{Schematic diagram of the $T=0$ phases of a strongly random 
electronic system in three dimensions described {\em e.g.\/}
by the Hamiltonian (\protect\ref{hamgeneral}). The average moment,
$\langle S_i \rangle$, when non-zero, varies randomly from site
to site. The conductivity is denoted by $\sigma$.}
\label{f1}
\end{figure}
\newpage
\begin{figure}
\epsfxsize=14in
\centerline{\epsffile{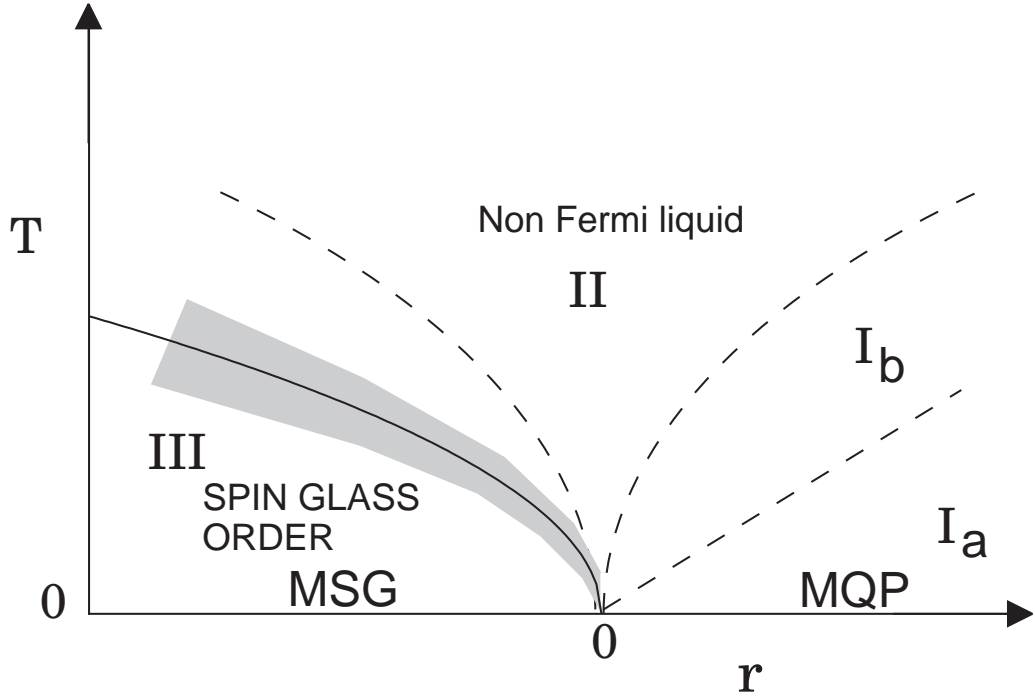}}
\vspace{0.5in}
\caption{Phase diagram of a metallic spin glass as a function of 
the ground state tuning parameter $r$ and temperature $T$. In the notation
of Fig~\protect\ref{f1}, the $T=0$ state is a MSG for $r<0$ and 
a MQP for $r>0$. 
The full line is the only thermodynamic phase transition, and is
at $r = r_c (T)$ or $T = T_c (r)$.
The quantum critical point is at $r=0$, $T=0$,
and is described by a continuum quantum field theory (CQFT). 
The dashed lines denote crossovers between different finite $T$ regions
of the CQFT: the low $T$ regions are Ia,Ib (on the paramagnetic side)
and III (on the ordered side), while the high $T$ region (II) displays
``non-Fermi liquid'' behavior.
The crossovers on either side of II, and the spin glass phase boundary
$T_c (r)$, all scale as $T \sim |r|^{z \nu/( 1 + \theta_u \nu)}$;
the boundary between Ia and Ib obeys $T \sim r^{z \nu}$. The mean field
values of these exponents are $z =4$, $\nu = 1/4$, and $\theta_u = 2$.
The shaded region has classical critical fluctuations described
by theories of the type discussed in Ref~\protect\cite{hertz}.
}
\label{f2}
\end{figure}
\newpage
\begin{figure}
\begin{picture}(500,230)
\put(20,210){\Large\bf LOW T REGION OF CQFT}
\put(50,160){\vector(1,0){400}}
\put(50,160){\line(0,1){20}}
\put(100,170){\Large Fermi liquid}
\put(230,160){\line(0,1){20}}
\put(300,170){\Large Critical}
\put(440,130){\Large $\omega$}
\put(227,140){\Large $r^{z \nu}$}
\put(50,140){\Large 0}
\put(20,80){\Large\bf HIGH T REGION OF CQFT}
\put(50,30){\vector(1,0){400}}
\put(50,30){\line(0,1){20}}
\put(60,40){\Large Quantum relaxational}
\put(230,30){\line(0,1){20}}
\put(300,40){\Large Critical}
\put(440,0){\Large $\omega$}
\put(227,10){\Large $u^{z \nu} T^{1+ \theta_u \nu}$}
\put(50,10){\Large 0}
\end{picture}
\vspace{0.5in}
\caption{Crossovers as a function of frequency, $\omega$, in the regions
of Fig~\protect\ref{f2}. The low $T$ region is on the paramagnetic
side ($r > 0$).}
\label{f3}
\end{figure}
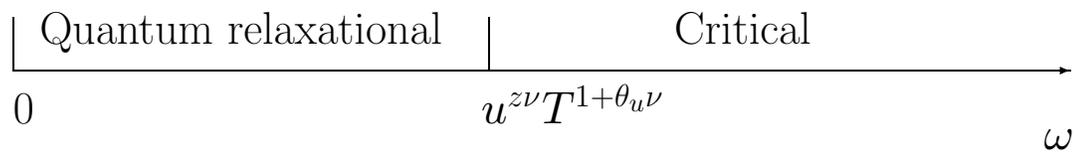
\end{document}